\documentclass[aps,prb,twocolumn,showpacs,preprintnumbers,amsmath,amssymb,superscriptaddress]{revtex4}%

\usepackage{graphicx}%
\usepackage{dcolumn}
\usepackage{amsmath}
\usepackage{color}
\usepackage{ulem}

\begin{document}

\title{High pressure magnetic state of MnP probed by means of muon-spin rotation}
\author{R.~Khasanov}
 \email{rustem.khasanov@psi.ch}
 \affiliation{Laboratory for Muon Spin Spectroscopy, Paul Scherrer Institut, 5232 Villigen PSI, Switzerland}
\author{A.~Amato}
 \affiliation{Laboratory for Muon Spin Spectroscopy, Paul Scherrer Institut, 5232 Villigen PSI, Switzerland}
\author{P.~Bonf\`{a}}
 \affiliation{Dipartimento di Fisica e Scienze della Terra e  Unit\`{a} CNISM di Parma, Universit\`{a} di Parma, 43124 Parma, Italy}
\author{Z.~Guguchia}
 \affiliation{Laboratory for Muon Spin Spectroscopy, Paul Scherrer Institut, 5232 Villigen PSI, Switzerland}
\author{H.~Luetkens }
 \affiliation{Laboratory for Muon Spin Spectroscopy, Paul Scherrer Institut, 5232 Villigen PSI, Switzerland}
\author{E.~Morenzoni}
 \affiliation{Laboratory for Muon Spin Spectroscopy, Paul Scherrer Institut, 5232 Villigen PSI, Switzerland}
\author{R.~De~Renzi}
 \affiliation{Dipartimento di Fisica e Scienze della Terra e  Unit\`{a} CNISM di Parma, Universit\`{a} di Parma, 43124 Parma, Italy}
\author{N.D.~Zhigadlo}
 \affiliation{Laboratory for Solid State Physics, ETH Zurich, 8093 Zurich, Switzerland,}
 \affiliation{Department of Chemistry and Biochemistry, University of Bern, 3012 Bern, Switzerland}

\begin{abstract}

We report a detailed $\mu$SR study of the pressure evolution of the magnetic order in the manganese based pnictide MnP, which has been recently found to undergo a superconducting transition under pressure once the magnetic ground state is suppressed. Using the muon as a volume sensitive local magnetic probe, we identify a ferromagnetic state as well as two incommensurate helical states (with propagation vectors ${\bf Q}$ aligned along the crystallographic $c-$ and $b-$directions, respectively) which transform into each other through first order phase transitions as a function of pressure and temperature. Our data appear to support that the magnetic state from which superconductivity develops at higher pressures is an incommensurate helical phase.

\end{abstract}


\maketitle

%

Recently, the binary pnictides CrAs and MnP  have attracted much interest due to discovery of superconductivity induced by hydrostatic pressure.\cite{Wu_NatConn_2014, Kotegawa_JPSJ_2014, Kotegawa_PRL_2015, Cheng_PRL_2015}  The helical magnet CrAs becomes superconducting for pressures exceeding $p\simeq 0.4$~GPa, whereas MnP possesses a critical pressure $p_c\simeq 8$~GPa at which magnetism disappears and  superconductivity sets in.
For CrAs, in the pressure range of $0.4\lesssim p\lesssim 0.7$~GPa, one observes a phase separation between  magnetic and paramagnetic volumes, with the latter becoming superconducting with a critical temperature $T_c\lesssim 2$~K. Above $0.7$~GPa the helical magnetic order vanishes and superconductivity below $T_c$ sets in within the whole sample volume.\cite{Khasanov_SciRep_2015} Note that in CrAs the single type of helical magnetic order remains unchanged as a function pressure.\cite{Khasanov_SciRep_2015}.
In comparison to CrAs, MnP possesses a more complicated phase diagram.\cite{Cheng_PRL_2015} At ambient pressure, MnP orders ferromagnetically at $T\simeq 290$~K  with the Mn magnetic moments aligned along the crystallographic $b-$direction (according to the crystallographic group $Pnma$ 62 with lattice constants $c>a>b$).\cite{Huber_PR_1964, Felcher_JAP_1966, Yamazaki_JPSJ_2014}  The ordered moment is $m\simeq 1.29$~$\mu_{\rm B}$ per Mn atom.\cite{Huber_PR_1964, Obara_JPSJ_1980} At lower temperatures ($T\lesssim 50$~K) another transition in a double-spiral helical structure is reported.\cite{Felcher_JAP_1966, Obara_JPSJ_1980, Forsyth_PPS_1966} In this helimagnetic state (helical$-c$ state, Ref.~\onlinecite{Matsuda_Arxiv_2016}) the Mn moments are rotated within the $ab$-plane (helical plane) with the propagation vector ${\bf Q}=(0,0,0.117)$.\cite{Felcher_JAP_1966} By increasing pressure the helical phase vanishes at $p\simeq 1-1.5$~GPa and a new magnetic phase emerges for $p\gtrsim 2$~GPa.\cite{Cheng_PRL_2015} In a narrow pressure region close to a critical pressure  $p_c$, at which the new magnetic phase disappears,  superconductivity is found below 1~K.
It was suggested, there is a quantum critical point at $p=p_c$ and that the quantum fluctuations persisting above $p_c$ give rise to occurrence of superconductivity.\cite{Cheng_PRL_2015}
Therefore, it is important to characterize the high-pressure (HP) magnetic state of MnP from which superconductivity emerges. At present, there are three reports based on the results of non-resonant x-ray, NMR, and neutron diffraction experiments pointing to a helical magnetic order.\cite{Matsuda_Arxiv_2016, Wang_Arxiv_2015, Fan_ScChin_2016} Note, that the NMR study does not allow one to identify the order, while the non-resonant x-ray and neutron diffraction experiments disagree on the type of HP helix. While the x-ray study points to a similar helical$-c$ structure  as the one at ambient pressure, the neutron diffraction study suggests that a conical or two-phase structure with ${\bf Q}\parallel b$ develops and it gradually changes to the so-called helical$-b$ structure with Mn moments rotating within the $ac$-plane.

\begin{figure}[htb]
\includegraphics[width=1.0\linewidth]{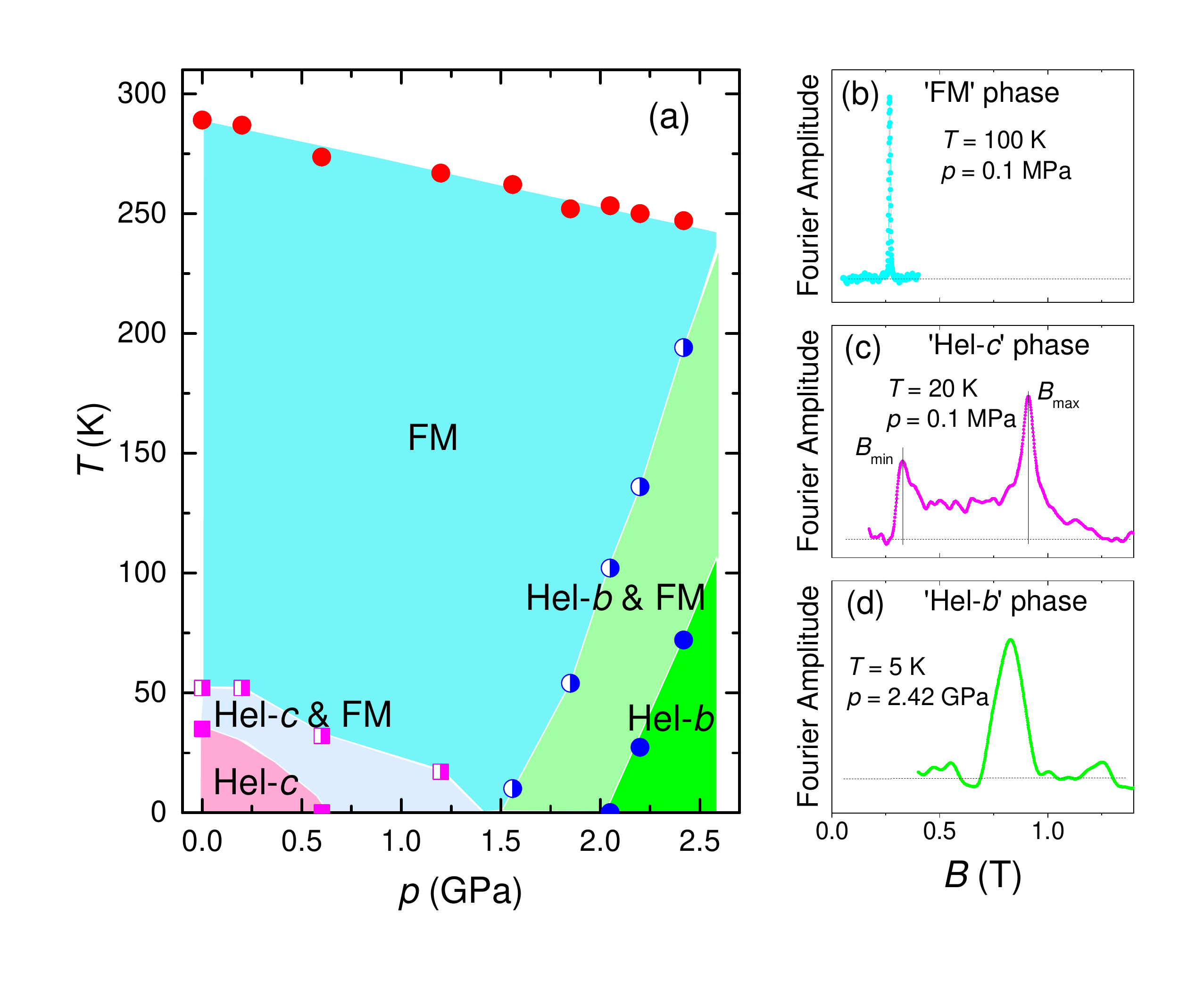}
%
\caption{(a) The $p-T$ phase diagram of MnP obtained from $\mu$SR experiments. FM, Hel$-c$, and Hel$-b$ denote the ferromagnetic, helical$-c$, and helical$-b$ magnetic orders, respectively (see text). The half filled and filled symbols correponds to the case when the Hel$-c$ (pink squares) or the Hel$-b$  (blue circles) phases occupy 10\% and 90\% of the sample volume, respectively. (b), (c), and (d) the magnetic field distribution obtained by means of ZF-$\mu$SR in the FM, Hel$-c$, and Hel$-b$ state, respectively. }
 \label{fig:Phase-Diagramm}
\end{figure}

In this paper we report on a detailed study of the evolution of magnetic properties of MnP as a function of temperature ($5\ {\rm K}\lesssim T\lesssim 300$~K) and pressure ($0.1{\rm ~MPa}\leq p\lesssim 2.4$~GPa) by means of muon-spin rotation ($\mu$SR). The resulting $p-T$ phase diagram (Fig.~\ref{fig:Phase-Diagramm}~a) consists of three areas corresponding to the ferromagnetic (FM), the helical$-c$ (Hel$-c$) and the helical$-b$ (Hel$-b$) orders with the corresponding field distributions [$P(B)$'s] presented in Figs.~\ref{fig:Phase-Diagramm}~b, c and d, respectively. The phases FM/Hel$-c$ as well as FM/Hel$-b$ are found to coexist within broad range of pressures and temperatures. Transitions from the high-temperature FM to the low-temperature low-pressure (LT-LP) Hel$-c$, or the low-temperature high-pressure (LT-HP) Hel$-b$ phases are first-order like. Our experiments appear to confirm that in MnP the high-pressure magnetic phase which is the precursor of the superconducting state is the incommensurate helical$-b$.

The magnetic response of MnP polycrystalline sample was studied in zero field (ZF) and weak transverse field (wTF) $\mu$SR experiments. A detailed description of $\mu$SR experiments under pressure, the construction of the pressure cell {\it etc.} are given in Ref.~\onlinecite{Khasanov_HPR_2016}. In the following we discuss the $\mu$SR data for the lowest (0.1~MPa) and the highest (2.42~GPa) pressures.

\begin{figure}[t]
\includegraphics[width=1.0\linewidth]{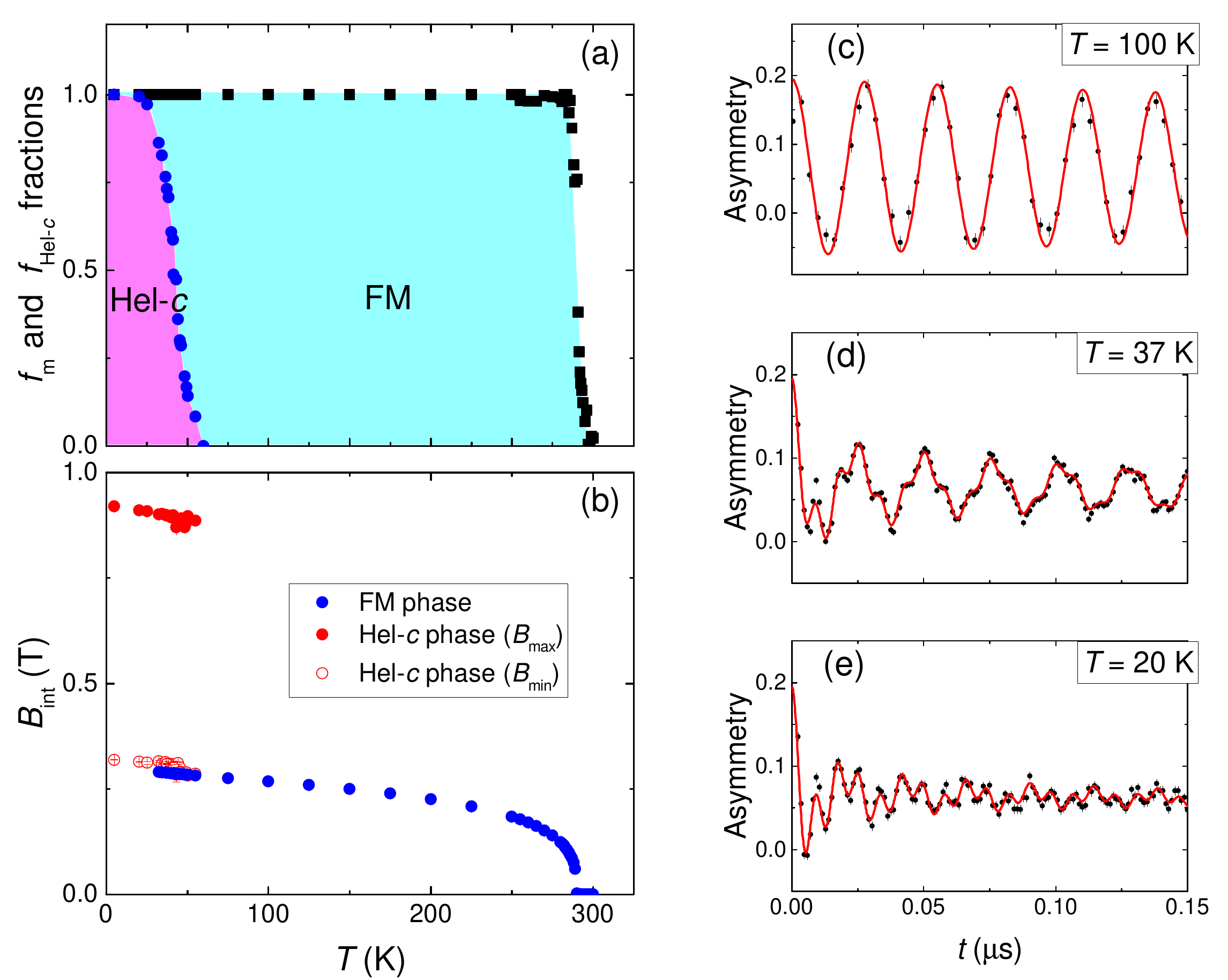}
%
\caption{The magnetic response of MnP at ambient pressure (0.1~MPa). (a) Temperature evolution of the total magnetic volume fraction ($f_{\rm m}$, black squares) and the helical$-c$ fraction ($f_{\rm Hel}$, blue circles). (b) Temperature dependence of internal fields within the FM and Hel$-c$ phases. (c), (d), and (e) ZF-$\mu$SR asymmetry spectra at $T=100$~K (FM phase), $T=37$~K (mixture of FM and Hel$-c$ phases), and $T=20$~K (Hel$-c$ phase). Solid red lines are fits as described in the Supplementary part. }
 \label{fig:ZP}
\end{figure}

At ambient pressure, transitions to two different magnetic states are  detected (Fig.~\ref{fig:ZP}~a). The first one at $T_{\rm FM}\simeq 290$~K corresponds to the ferromagnetic order.\cite{Huber_PR_1964, Felcher_JAP_1966}  The spontaneous muon-spin precession is clearly observed in the ZF $\mu$SR asymmetry spectra  [$A(t)$, Fig.~\ref{fig:ZP}~c]. The oscillatory part of $A(t)$ is well fitted by an exponentially decaying cosine function with a zero initial phase (see the Supplementary part), thus evidencing that the magnetic order is commensurate.\cite{Yaouanc_book_11} In the
LT-LP region, $T\lesssim25$~K, the oscillatory part of $A(t)$ is accurately described by a field distribution characterized by a minimum ($B_{min}$) and a maximum ($B_{max}$) cutoff fields (see Figs.~\ref{fig:Phase-Diagramm}~c and \ref{fig:ZP}~e), which is consistent with the incommensurate helimagnetic order.\cite{Schenk_PRB_2001}
In a broad range of temperatures ($30\ {\rm K}\lesssim T\lesssim 50$~K) both orders are detected simultaneously (Fig.~\ref{fig:ZP}~d). Being a volume sensitive local probe  technique, $\mu$SR allows to follow the temperature evolution of both magnetic phases as demonstrated in Fig.~\ref{fig:ZP}~a. The corresponding internal fields in the FM ($B_{int}$) and Hel$-c$ ($B_{min}$, $B_{max}$) phases are presented in Fig.~\ref{fig:ZP}~b. The internal magnetic field in the FM phase ({\it i.e.}, the FM magnetic order parameter) decreases with increasing temperature and vanishes at $T_{\rm FM}$. Analyzing the data with a power law
\begin{equation}
B_{int}(T)=B_{int}(T=0)[1-(T/T_{\rm FM})^\alpha]^\beta
 \label{eq:Bint}
\end{equation}
yields $T_{\rm FM}=289.0(2)$~K, $B_{int}(T=0)=0.2925(2)$~T, $\alpha=1.16(1)$ and $\beta=0.252(3)$. The value of $\beta$ lies quite close to the critical exponent $\beta\simeq 1/3$ expected for a second order phase transition in a 3D magnetic system. \cite{Jongh_AdvPhys_1974}
In contrast, both $B_{min}$ and $B_{max}$, which are measures of the magnetic order parameter of the Hel$-c$ state, abruptly drop to zero at the phase transition (Fig.~\ref{fig:ZP}~b). This, together with the coexistence of the FM and Hel$-c$ phases  in a relatively large temperature region of $\sim20$~K indicates the first order character of the transition.

\begin{figure}[htb]
\includegraphics[width=1.0\linewidth]{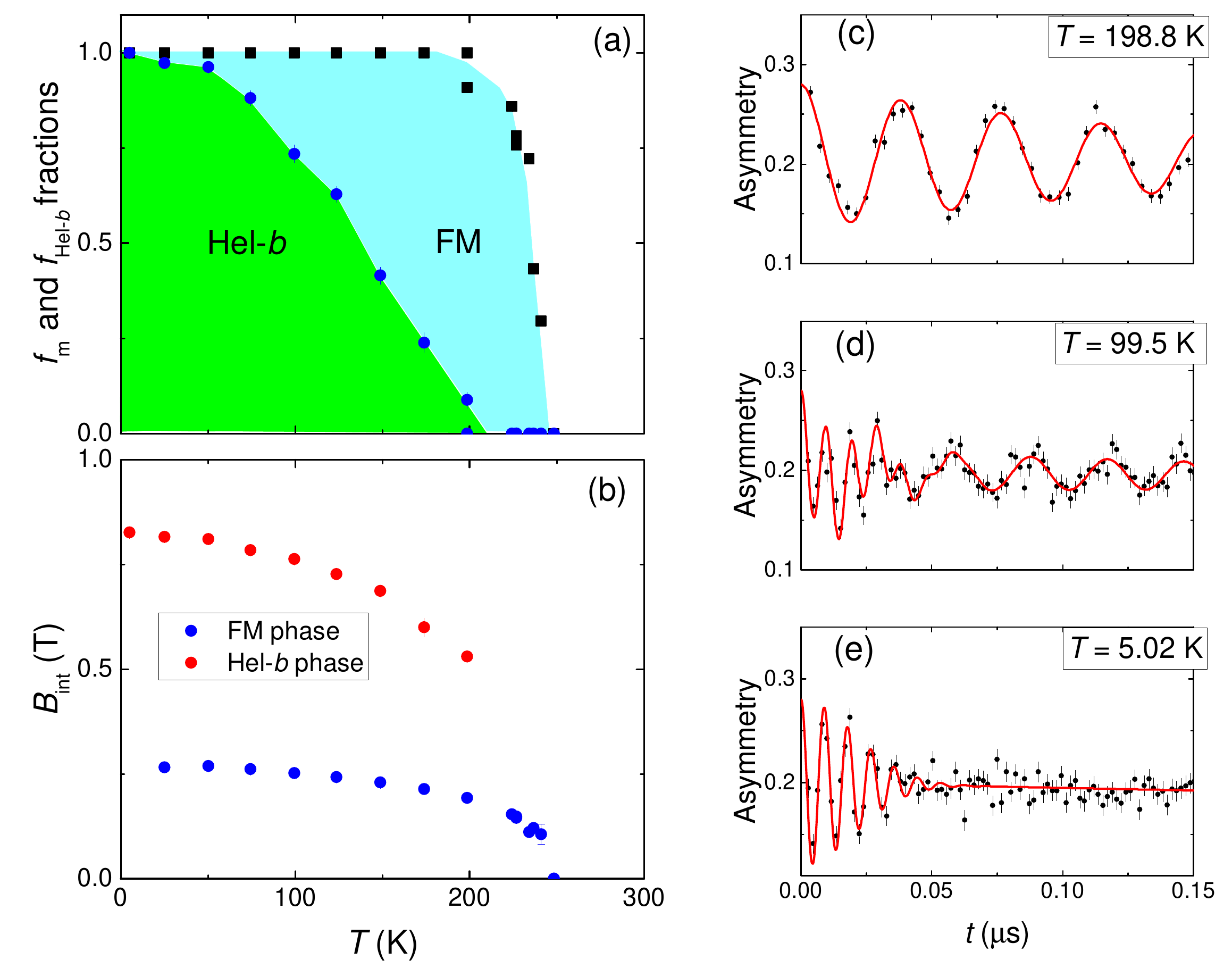}
 \vspace{-0.5cm}
\caption{Magnetic response of MnP at $p=2.42$~GPa. (a) Temperature evolution of the total magnetic volume fraction ($f_{\rm m}$, black squares) and the LT-HP (Hel$-b$) fraction ($f_{{\rm Hel}-b}$, blue circles). (b) Temperature dependence of internal fields within the FM and LT-HP phases. (c), (d), and (e) ZF-$\mu$SR asymmetry spectra at $T=198.8$~K (FM phase), $T=99.5$~K (mixture of FM and LT-HP phases), and $T=5.02$~K (LT-HP phase). Solid red lines are fits as described in the Supplementary part.}
 \label{fig:HP}
\end{figure}

At $p=2.42$~GPa, the transition into the FM state is clearly detected  via the observation of a spontaneous muon-spin precession frequency for temperatures below 250~K. Upon decreasing the temperature, another phase, with an internal field approximately 2.5 times higher, starts to develop for $T\lesssim 200$~K and occupies the full sample volume for $T\lesssim 50$~K (Figs.~\ref{fig:HP}a and b). This phase  corresponds to the `unknown antiferromagnetic' phase which was first reported in Ref.~\onlinecite{Cheng_PRL_2015}.
A preliminary analysis reveals that the oscillatory part of the $\mu$SR signal
in the LT-HP phase is well fitted by the Gaussian decaying cosine function with zero initial phase. The temperature dependences of the magnetic volume fractions and the internals fields obtained for the FM and LT-HP phases are shown in Figs.~\ref{fig:HP}~a and b. The asymmetry spectra taken at $T\simeq 200$~K (FM phase), $\simeq 100$~K (mixture of FM and LT-HP magnetic phases), and $\simeq 5$~K (LT-HP magnetic phase) are shown in panel c, d, and e, respectively.

At first glance the ambient pressure and $p=2.42$~GPa data presented in Figs.~\ref{fig:ZP} and \ref{fig:HP} look similar. Indeed, (i) in both cases the ferromagnetic phase observed at high temperatures is fully replaced by a LT phase. (ii) within a broad range of temperatures the ferromagnetic and LT phases coexist. (iii) the appearance of the LT phase is always associated with an abrupt change of the internal field (magnetic order parameter), {\it i.e.} there is a first order transition between the FM and both LT phases.

There is, however, an important difference between them. At ambient pressure the Mn moments form the so-called `double helical' structure with the propagation vector ${\bf Q}=(0,0,0.117)$. Considering 4 Mn atoms per unit cell, the term `double helical' means that the Mn spins are coupled in pairs (Mn1/Mn4 and Mn2/Mn3), and rotate with a constant phase difference between the different pairs along ${\bf Q}$ (see Ref.~\onlinecite{Felcher_JAP_1966}).  Such an incommensurate magnetic structure leads to a field distribution given by: \cite{Schenk_PRB_2001}
\begin{equation}
P(B)=\frac{2}{\pi}\frac{B}{\sqrt{(B^2-B^2_{min})(B^2_{max}-B^2)}}
 \label{eq:helical_PB}
\end{equation}
and is characterized by two peaks at a minimum ($B_{min}$) and a maximum($B_{max}$) cutoff field (Fig.~\ref{fig:Phase-Diagramm}~c).

The magnetic field distribution in the LT-HP phase is, however, different. It is characterized by a {\it single} symmetric line (Fig.~\ref{fig:Phase-Diagramm}~d), though with a very broad width.
At first glance such $P(B)$ is inconsistent with the helical order. However, as shown below, the unique situation realized in MnP may lead to  the formation of a quasi-single peak structure. In short, in MnP muons stop at four well defined interstitial lattice sites within the unit cell. The $\mu$SR asymmetry spectra consists, therefore, of four contributions with each of them characterized by its own $B_{min}$ and $B_{max}$ fields and corresponding $P(B)$'s described by Eq.~(\ref{eq:helical_PB}). Under certain conditions which, as shown in the Supplementary part, are fulfilled in MnP for a  helical$-b$ magnetic order, the sum of the four asymmetric $P(B)$ distributions results in a single broad symmetric line.

The exact spin arrangement of the LT-HP magnetic phase is currently under debate. It is also not clear whether this phase has a non-zero net magnetic moment. Following Fig.~S2 in the Supplementary part of Ref.~\onlinecite{Cheng_PRL_2015} the ferromagnetic component of the LT-HP phase might be quite substantial and correspond to $\sim 5-20$\% of that in the FM phase.
The authors of Ref.~\onlinecite{Matsuda_Arxiv_2016} have also suggested a conical magnetic structure for $p\gtrsim 1.5$~GPa with a ferromagnetic component gradually decreasing with decreasing temperature and increasing pressure.
We checked for a possible ferromagnetic component by studying the  pressure cell response in a wTF $\mu$SR experiments. Note that the spins of the  muons stopping in the pressure cell containing the magnetic sample will undergo a precession around the vector sum of the weak externally applied field and the straight fields induced by the sample. If these straight fields are sizable and inhomogeneous, as expected around a ferromagnetic sample, the $\mu$SR signal of the muons stopping in the cell will exhibit a relaxation due to dephasing.\cite{Maisuradze_PRB_2011}
\begin{figure}[t]
\includegraphics[width=0.8\linewidth]{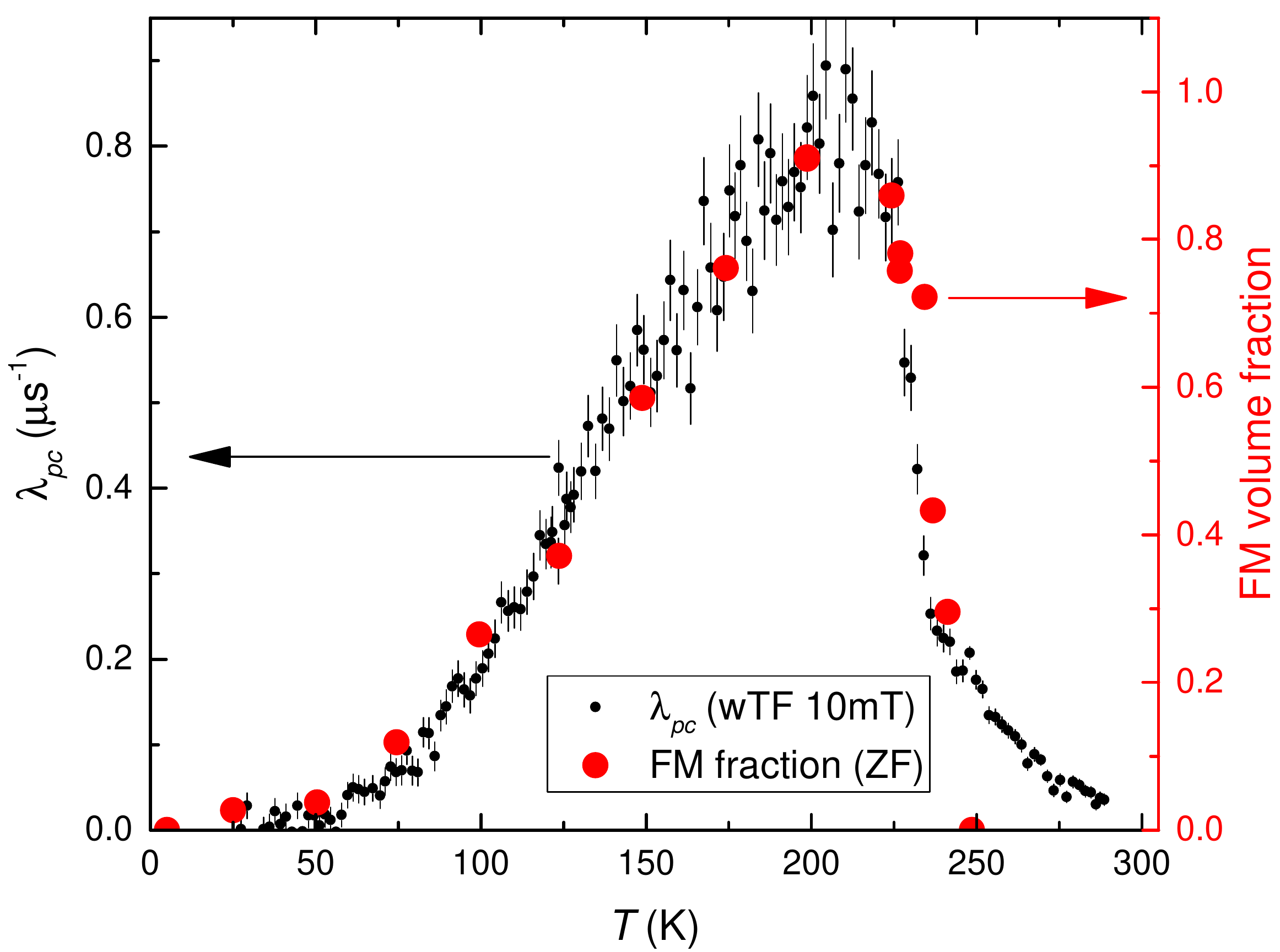}
%
\caption{Temperature dependences of $\lambda_{pc}$ obtained in 10~mT wTF $\mu$SR experiments (black dots) and the FM volume fraction $f_{\rm FM}$ measured in ZF-$\mu$SR experiments (red circles) at $p=2.42$~GPa.  }
 \label{fig:WTF-lambda_PC}
\end{figure}
Figure~\ref{fig:WTF-lambda_PC} shows the temperature dependence of an additional exponential relaxation of the spin polarization for the muons stopping in the pressure cell ($\lambda_{pc}$). The magnetic field $\mu_0H=10$~mT, transverse to the initial muon-spin polarization, was applied at $T\simeq300$~K (above the magnetic transition). For comparison we also plot in this graph the temperature evolution of the ferromagnetic volume fraction ($f_{\rm FM}=f_{\rm m}-f_{{\rm hel}-b}$) as obtained in ZF $\mu$SR experiments (Fig.~\ref{fig:HP}a). Below $T\simeq 250$~K $\lambda_{pc}$ follows {\it exactly} $f_{\rm FM}$. With decreasing temperature both quantities first increase, thus showing the formation of the FM phase. Both, $\lambda_{pc}$ and $f_{\rm FM}$, saturate at $T\simeq 200$~K, and then decrease almost linearly with temperature decrease from $\simeq200$~K down to $\simeq 50$~K. Below 50~K, where the FM phase completely vanishes, the signal of the pressure cell becomes unaffected by the sample.  This proves that the magnetic order induced by high pressure in MnP sample possesses no sizable ferromagnetic component and allows one to exclude from the consideration the conical magnetic structure.

\begin{figure}[t]
\includegraphics[width=0.8\linewidth]{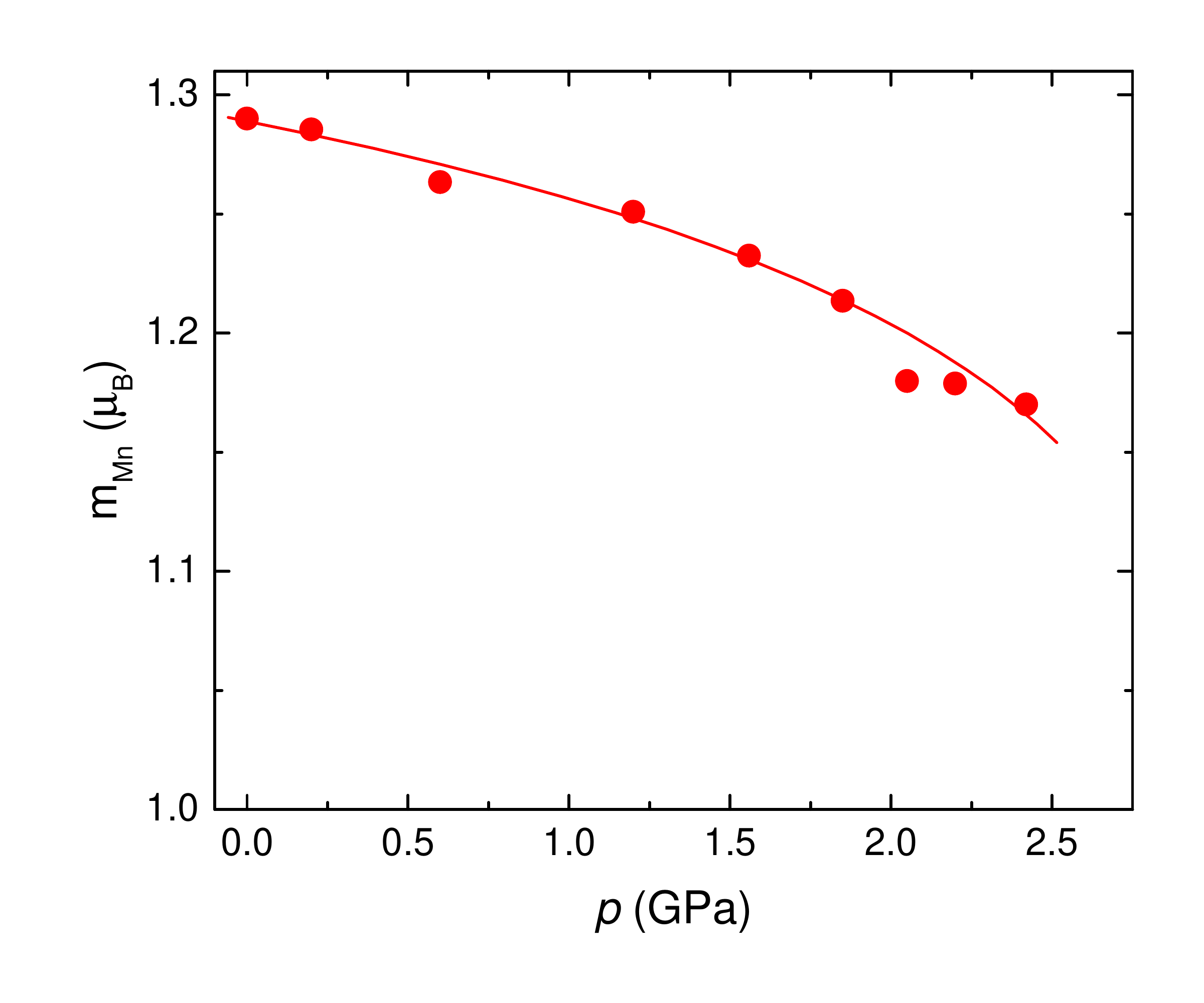}
%
\caption{ Pressure dependence of the ordered moment $m$ in the FM state. The line is the guide for the eye.
}
 \label{fig:Moment_vs_pressure}
\end{figure}

Additional experiments performed in the intermediate range of pressures ($p=0.2$, 0.6, 1.2, 1.56, 1.8, 2.05, and 2.20~GPa) allow us to construct the  full $p-T$ phase diagram as shown in Fig.~\ref{fig:Phase-Diagramm}a. It resembles the one reported in Ref.~\onlinecite{Cheng_PRL_2015}, with, however, some important differences. Firstly, the transition temperature of the FM state changes quite monotonically with pressure without a pronounced ${\rm d} T_c/{\rm d} p$ slope change at $p\sim 2$~GPa. The slope ${\rm d} T_c/{\rm d} p = -17.5(9)$~K/GPa is in agreement with the results of Banus \cite{Banus_JSSCh_1972} reporting ${\rm d} T_c/{\rm d} p = -18.5$~K/GPa for pressures ranging $0\leq p \lesssim 2.0$~GPa.
Secondly, there is a broad temperature range where FM and Hel$-c$ (LT-LP), and FM and Hel$-b$ (LT-HP) orders coexist. We do not detect, however, any coexistence of Hel$-c$ and Hel$-b$ magnetic orders.

All above conclusions do not require any particular modeling and are obvious already from the $\mu$SR raw data. To obtain more quantitative information calculations of local fields ($B_{\rm loc}$) at the muon stopping sites were carried out. The muon site was determined based on the density-functional theory (DFT) approach, which was shown to accurately determine the muon sites in different materials as {\it e.g.} wide-gap semiconductors, insulating systems or cuprate and iron-based high-$T_c$ superconductors. \cite{Bernardini_PRB_2013, Moller_PRB_2013, Blundell_PRB_2013, Moller_PhScr_2013, Amato_PRB_2014, Martin_Arxiv_2016} There are four
equivalent minima in the unit cell corresponding to the 4c [$(x,1/4,z)$, $x_\mu=0.103$, $z_\mu=0.921$] Wyckoff position (see the Supplementary part).
The calculations were started from the known FM and Hel$-c$ magnetic structures at ambient pressure (see Refs.~\onlinecite{Huber_PR_1964, Felcher_JAP_1966, Obara_JPSJ_1980, Forsyth_PPS_1966}). From this first step one could determine the coupling contact constant $A_{\rm cont}\simeq -0.447$~${\rm T}/\mu_{\rm B}$. As a next step, by assuming $A_{\rm cont}$ and the relative positions of Mn ions and muon within the unit cell are pressure independent, as well as accounting for the pressure reduction of lattice constants,\cite{Wang_Arxiv_2015} the pressure dependence of the ordered moment $m$ in the FM state was calculated (Fig.~\ref{fig:Moment_vs_pressure}). The internal field at $T=0$ was obtained from the fit of $B_{int}(T)$ by means of Eq.~\ref{eq:Bint} measured at each particular pressure . We have also checked for consistency of the LT-HP magnetic phase with 3 possible collinear antiferromagnetic structures reported in Ref.~\onlinecite{Gercsi_PRB_2010} and the helical structure proposed in Ref.~\onlinecite{Wang_Arxiv_2015}. The analysis reveals that none of them are consistent with the experimental data (see the Supplementary part).

To test whether the helical$-b$ structure proposed in Ref.~\onlinecite{Matsuda_Arxiv_2016} is compatible with our data, we started first by  calculating the $B_{min}$ and $B_{max}$ fields for each particular muon site in the range of $-0.14\leq \delta \leq 0.14$ and $-0.16\leq e \leq 0.16$. Here $\delta$ is the component of the propagation vector ${\bf Q}=(0,\delta,0)$
(the positive and the negative sign correspond to a right and a left-handed helix, respectively) and $e=(m_a-m_c)/(m_a+m_c)$ is the eccentricity of the elliptical helical$-b$ structure ($m_a$ and $m_c$ are components of $m$ along the $a-$ and $c-$axis, respectively).
With such determined sets of $B_{min}$'s and $B_{max}$'s the experimental asymmetry spectra were fitted. The correlation plots $\chi^2(\delta,e)$ and $m(\delta,e)$ allow us to check for the range of consistency of helical$-b$ structure with the experimental data and to obtain the value of the average moment $m$. With $\delta=\pm 0.09$ (Ref.~\onlinecite{Matsuda_Arxiv_2016}), $e$ and $m$ were found to range $-0.025\lesssim e \lesssim 0.05$ and $1.175\ \mu_{\rm B}\lesssim m \lesssim 1.190\ \mu_{\rm B}$, respectively. Note that the value of $m$ in the helical$-b$ state is close to that determined in the ferromagnetic state $m_{\rm FM}=1.17$~$\mu_{\rm B}$. The details of calculations are summarized in the Supplementary Information.

To conclude, the muon-spin rotation measurements of MnP under the pressure up to $\simeq2.4$~GPa were carried out. The ferromagnetic ordering temperature and the value of the ordered Mn moments decrease with the pressure increase from $T_{\rm FM}=289.0(2)$~K and $m=1.29$~$\mu_{\rm B}$ at $p=0.1$~MPa, to  $T_{\rm FM}=242.0(2.2)$~K and $m=1.17$~$\mu_{\rm B}$ at $p=2.42$~GPa, respectively. For pressures in the region $0.1{\rm ~MPa}\leq p \lesssim 1.2$~GPa the ferromagnetic  and the LT-HP helical$-c$  types of magnetic order were clearly detected. The helical$-c$ order becomes completely suppressed for pressures exceeding 1.5~GPa. Above this pressure the third magnetic phase, the helical$-b$ phase,\cite{Matsuda_Arxiv_2016}  starts to grow. The transition temperature and the volume fraction of the helical$-b$ phase increase continuously with increasing pressure. At $p=2.42$~Ga the helical$-b$ phase was clearly detected up to $T\simeq200$~K. Pairs of phases, FM/Hel$-c$ and FM/Hel$-b$, coexist within broad range of pressures and temperatures. Transitions from the HT FM to the LT-LP helical$-c$, or the LT-HP helical$-b$ phases are first-order like.

The work was fully performed at the Swiss Muon Source (S$\mu$S), PSI, Villigen. The work of ZG was supported  by the Swiss National Science Foundation (SNF-Grant 200021-149486). PB thanks the computing resources provided by CINECA within the Scientific Computational Project CINECA ISCRA Class C (Award HP10C5EHG5, 2015) and STFC's Scientific Computing Department. The work PB and RdR was supported by the grants from MUON JRA of EU FP7 NMI3, under grant agreement 226507.

\clearpage

\section*{Supplementary Information}

\setcounter{equation}{0} \setcounter{figure}{0}

\renewcommand{\theequation}{S\arabic{equation}}
\renewcommand{\thefigure}{S\arabic{figure}}

\subsection{Sample preparation}

The manganese phosphide (MnP) polycrystalline sample was synthesized by using a high-pressure furnace. Overall details of the sample cell assembly and high-pressure synthesis process can be found in Ref.~\onlinecite{Zhigadlo_PRB_2012_sup}. The mixture of manganese powder (99.99\%) and red phosphorus powder (99.999\%) in a molar ratio 1:1 was enclosed in a boron-nitride crucible and placed inside a pyrophyllite cube with a graphite heater. All the preparatory steps were done in a glove box under argon atmosphere. In a typical run, the sample was compressed up to 1~GPa at room temperature. While keeping pressure constant, the temperature was ramped up within 3~h to the maximum value of 1200~$^{\rm o}$C, kept stable for 1~h, then cooled to 950~$^{\rm o}$C in 14~h and finally quenched to the room temperature. Afterwards, the pressure was released and the final solid product removed. Subsequently recorded x-ray powder diffraction patterns showed no secondary phases.

\subsection{Experimental Techniques}

\subsubsection{Pressure Cell}

The pressure was generated in a double-wall piston-cylinder type of cell made of MP35N alloy. As a pressure transmitting medium 7373 Daphne oil was used. The pressure was measured in situ by monitoring the pressure shift of the superconducting transition temperature of In. The details of the experimental setup for conducting $\mu$SR under pressure experiments are given in Ref.~\onlinecite{Khasanov_HPR_2016_sup}.

\subsubsection{Muon-spin rotation}

$\mu$SR measurements at zero field (ZF) and field applied transverse to the initial muon-spin polarization were  performed at the $\pi$M3 and $\mu$E1 beamlines (Paul Scherrer Institute, Villigen, Switzerland), by using the dedicated GPS and GPD\cite{Khasanov_HPR_2016_sup} spectrometers, respectively. At the GPS spectrometer, equipped with continues flow $^4$He cryostat, ZF experiments at ambient pressure and down to temperatures $1.6$~K  were carried out. At the GPD spectrometer, equipped with continuous flow $^4$He cryostat (base temperature $\simeq 2.2$~K), ZF, and 10~mT weak transverse field (wTF) $\mu$SR experiments under pressure up to $\sim$2.4~GPa were conducted. All ZF experiments were performed by stabilizing the temperature prior to recording the muon-time spectra. In wTF experiments the temperature was swept up with the rate $\simeq0.2$~K/min. The data were collected continuously. Each muon-time spectra was recorded during approximately 5 minutes.

\subsection{ZF $\mu$SR data analysis procedure}

At ambient pressure MnP is either in the paramagnetic state (PM, $T\gtrsim 290$~K) or exhibits ferromagnetic (FM, $50\lesssim T \lesssim 290$~K) or helical-c (Hel$-c$, $T\lesssim 50$~K) magnetic order.

In the PM state the muon-spin polarization is well described by the single exponential decay function:
\begin{equation}
P^{\rm PM}(t)=e^{-\lambda t}.
 \label{eq:P_PM}
\end{equation}
Here $\lambda$ is the exponential relaxation rate.

In the FM state the muon-spin polarization follows:
\begin{equation}
P^{\rm FM}(t)=\frac{2}{3}e^{-\lambda_T t} \cos(\gamma_\mu
B_{int}t) + \frac{1}{3}\;e^{-\lambda_L t},
 \label{eq:P_FM}
\end{equation}
where $B_{int}$ is the internal field on the muon stopping site, $\gamma_{\mu}= 2\pi\;135.5$~MHz/T is the muon gyromagnetic ratio, and $\lambda_T$ and $\lambda_L$ are the transverse and the longitudinal exponential relaxation rates, respectively. The occurrence of 2/3 oscillating and 1/3 non oscillating $\mu$SR signal fractions originates from the spatial averaging in powder samples, where 2/3 of the magnetic field components are perpendicular to the muon-spin and cause a precession, while the 1/3 longitudinal field components do not.

In the helical$-c$ state the magnetic field distribution is characterized by two peaks due to the minimum ($B_{min}$) and maximum($B_{max}$) cutoff fields (see the Fig.~1c in the main text). Following Ref.~\onlinecite{Amato_PRB_2014_sup} the muon-spin polarization for the helical type of magnetic order is well described as:
\begin{eqnarray}
P^{\rm Hel}(t)&=&\frac{2}{3}e^{-\lambda_T t} J_0(\gamma_\mu \Delta B t)\cos(\gamma_\mu B_{av}t) +  \label{eq:P_Hel} \\
&&+\frac{1}{3}\;e^{-\lambda_L t}. \nonumber
 \end{eqnarray}
Here $J_0$ is the zeroth order Bessel function of the first kind, and the widths of the distribution $\Delta B$ and the average field $B_{av}$ are $\Delta B= (B_{max}-B_{min})/2$ and $B_{av}=(B_{max}+B_{min})/2$, respectively.

Considering three above described states the ZF-$\mu$SR asymmetry spectra of MnP at ambient pressure were analyzed as:
\begin{eqnarray}
A_s(t)&=&A_s(0)\{(1-f_m) P^{\rm PM}(t) +    \label{eq:Asymmetry_Separation} \\
&&+f_m [(1-f_{\rm Hel}) P^{\rm FM}(t) + f_{\rm Hel} P^{\rm Hel}(t)] \}. \nonumber
\end{eqnarray}
Here $A_s(0)$ is the initial asymmetry at $t=0$, and $f_m$ and $f_{\rm Hel}$ are the total magnetic volume fraction and the volume fraction of the helical phase, respectively.

In pressure experiments a large fraction of the muons, roughly 50\%, stops in the pressure cell walls. The fit function consists of the "sample" and the background (pressure cell) contributions and is described as:
\begin{equation}
A(t)=A_s(0) P_s(t)+ A_{pc}(0) P_{pc}(t).
 \label{eq:ZF_Asymmetry_PC}
\end{equation}
Here $A_{s}(0)$ and $A_{pc}(0)$ are the initial asymmetries and $P_{s}(t)$ and $P_{pc}(t)$ are the muon-spin polarizations belonging to the sample and the pressure cell, respectively. The polarization of the pressure cell is generally studied in separated set of experiments.\cite{Khasanov_HPR_2016_sup}

For pressures ranging from 0.1~MPa to $\simeq 1.2$~GPa, where the MnP sample obeys transition from the FM to the Hel$-c$ state (see Fig.~\ref{fig:Phase-Diagramm}~a), the sample response was analyzed by using Eq.~\ref{eq:Asymmetry_Separation} with the individual components described by Eqs.~\ref{eq:P_PM}, \ref{eq:P_FM}, and \ref{eq:P_Hel}, respectively.

For pressures above $1.5$~GPa the response of the sample in the Hel$-b$ phase, was found to be well described as:
\begin{equation}
P^{\rm FM}(t)=\frac{2}{3}e^{-\sigma_T^2 t^2/2} \cos(\gamma_\mu
B_{int}t) + \frac{1}{3}\;e^{-\lambda_L t},
 \label{eq:P_AFM}
\end{equation}
where $\sigma_T$ is the transverse Gaussian relaxation rate. The average internal field in the Hel$-b$ phase is approximately 2.5 times higher than that in the FM phase ($B_{int}^{{\rm Hel}-b}\simeq 2.5 B_{int}^{\rm FM}$, Figs.~\ref{fig:HP}~b, c and e in the main text). Such big difference allows us easy to distinguish between contributions of FM and Hel$-b$ phases in the $\mu$SR asymmetry spectra. Fits were made by using Eq.~\ref{eq:Asymmetry_Separation} with the 'Hel` part replaced by the 'Hel$-b$` one and the individual components were described by Eqs.~\ref{eq:P_PM}, \ref{eq:P_FM}, and \ref{eq:P_AFM}, respectively.

\subsection{wTF $\mu$SR data analysis procedure}

\begin{figure}[htb]
\includegraphics[width=0.9\linewidth]{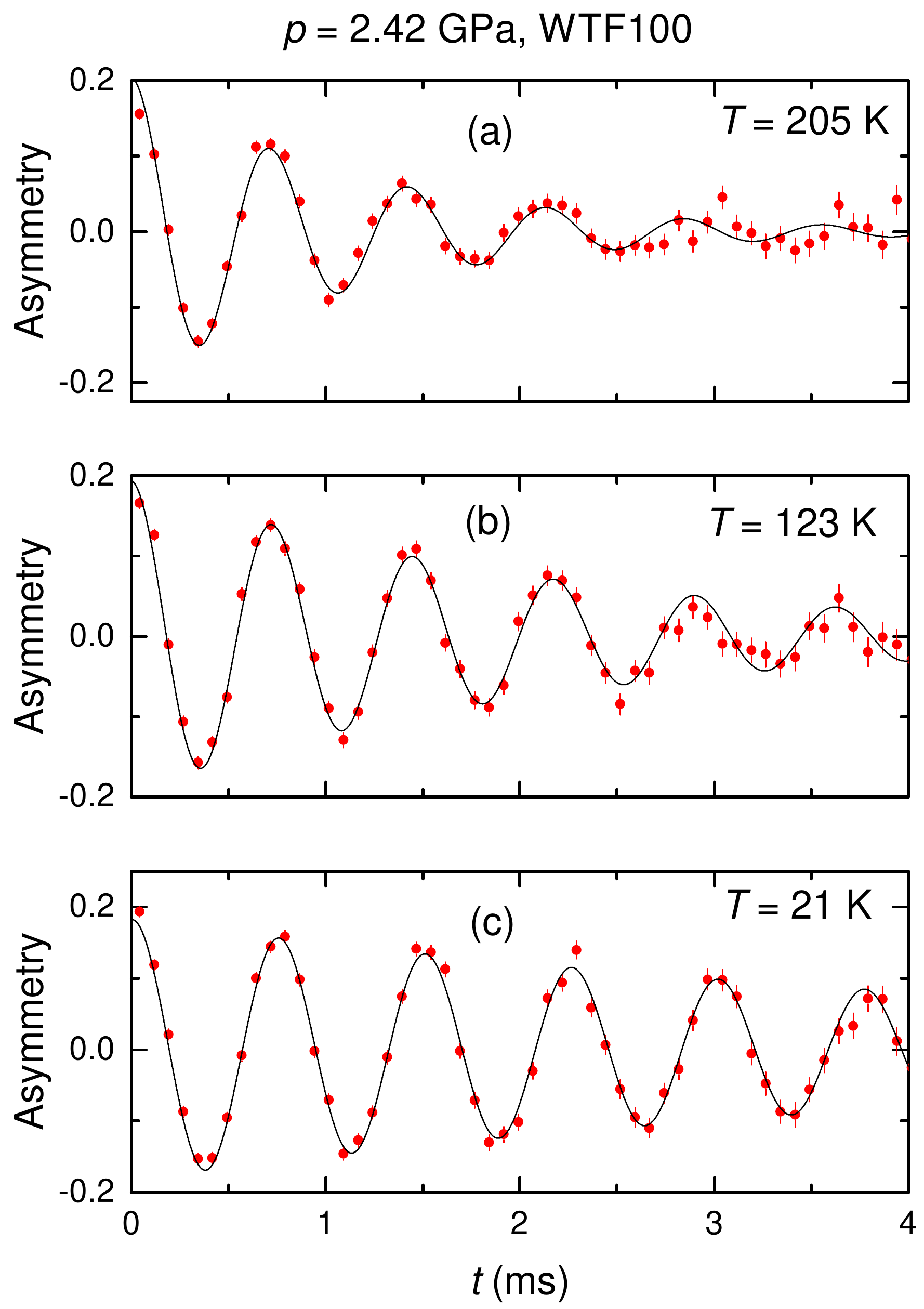}
%
\caption{ The pressure cell response in wTF $\mu$SR experiment at $p=2.42$~GPa, $\mu_0H=10$~mT and temperatures $T=205$~K (a), 123~K (b), and  21~K (c).  The solid lines are fits by using Eq.~\ref{eq:PC_TF}. }
 \label{fig:WTF-GPD_PC}
\end{figure}

Samples with a strong magnetization placed in a pressure cell may induce a magnetic field in the space around the sample. Typical examples of such samples are superconductors (strong diamagnets), superparamagnets, and ferro- or ferrimagnets. Thus, muons stopping in a pressure cell containing the sample will undergo precession in the vector sum of the applied field and the field induced
by the sample. This spatially inhomogeneous field leads to an additional depolarization of the muon spin polarization, which depends on the applied field and the induced field together with the spatial stopping distribution of the muons.\cite{Maisuradze_PRB_2011_sup}

Figure~\ref{fig:WTF-GPD_PC} shows weak-transverse field (wTF) $\mu$SR asymmetry spectra taken at $p=2.42$~GPa and $T=205$~K (panel a), 123~K (panel b) and 21~K (panel c). The magnetic field $\mu_0H=10$~mT was applied at $T\simeq300$~K (above the magnetic transition). Note that signals in Fig.~\ref{fig:WTF-GPD_PC} are mostly determined by the muons stopped in the pressure cell walls. The sample contribution, with internal fields $B_{int}\simeq 250$~mT in the FM state or $\simeq 800$~mT in the Hel$-b$ state, is suppressed with the proper selection of 'binning` [data points were grouped together with the corresponding final step of $\simeq 70$~ns ($\simeq 1$~mT)]. Obviously, at $T=205$~K there is a strong influence of the sample magnetization on the pressure cell response. It becomes much weaker at $T=123$~K and vanishes at $T=21$~K. Note that at $T=21$~K the signal corresponds to the undistorted pressure cell response.

The data presented in Fig.~\ref{fig:WTF-GPD_PC} were analyzed by using the following functional form:
\begin{equation}
A_{pc}(t)=A_{pc}(0)\; e^{-\lambda_{ pc}t}\; e^{-\sigma_{pc}^2t^2/2}\; \cos(\gamma_\mu B_{pc}t+\phi)
 \label{eq:PC_TF}
\end{equation}
Here $\sigma_{pc}$ is the Gaussian relaxation rate caused by nuclear moments and $\lambda_{pc}$ is the exponential relaxation reflecting the influence of  the sample on the pressure cell.

\subsection{Internal fields on muon stopping sites}

\subsubsection{MnP unit cell}

\begin{figure}[htb]
\includegraphics[width=0.8\linewidth]{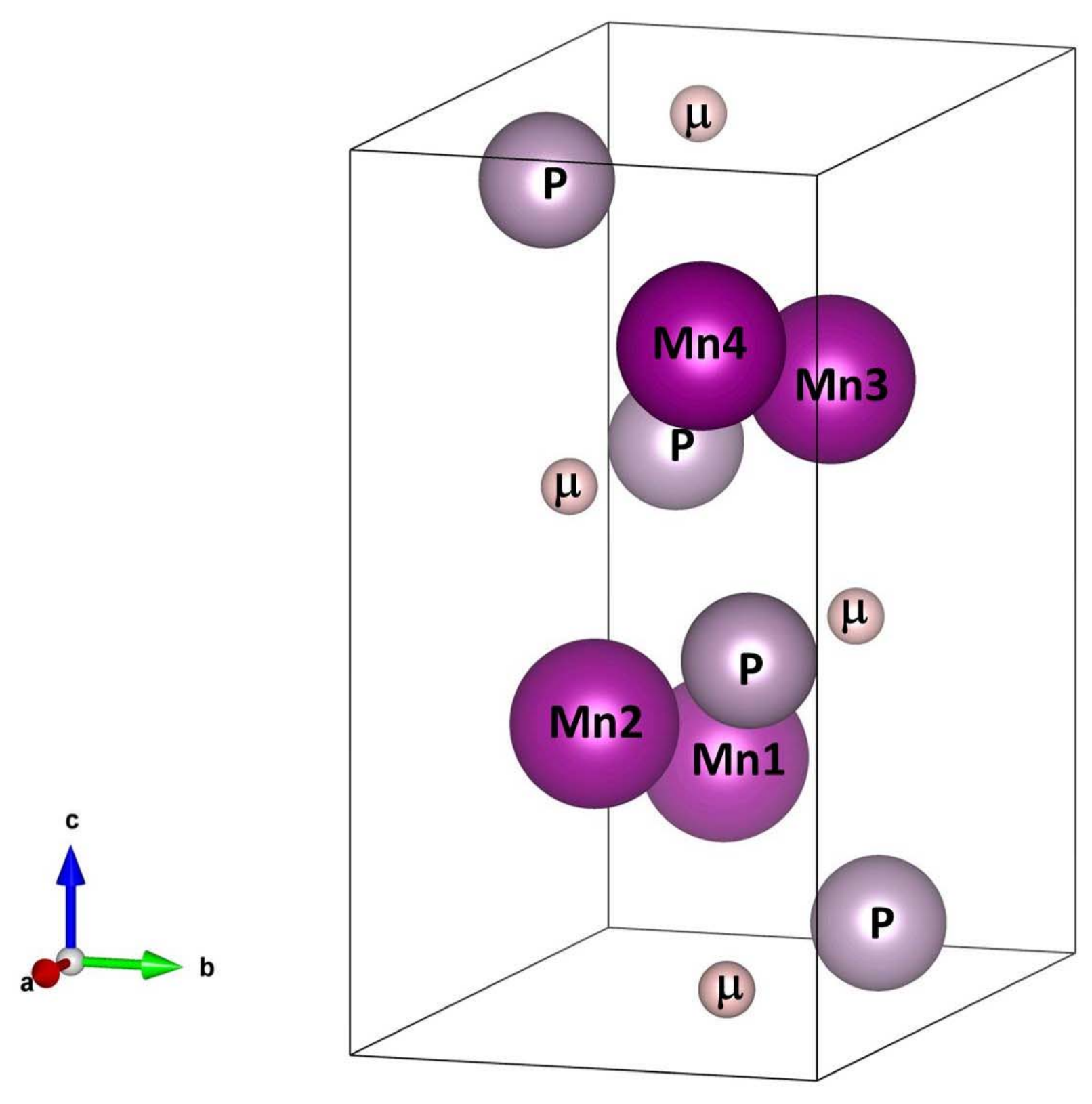}
%
\caption{The orthorhombic crystal structure of MnP ($Pnma$, 62). The muon stopping positions were obtained by $ab-initio$ calculations (see text for details). The structure  was visualized by using VESTA.\cite{Vesta}
}
 \label{fig:MnP_Unit-cell}
\end{figure}

The orthorhombic crystal structure of MnP ($Pnma$, 62) is shown in Fig.~\ref{fig:MnP_Unit-cell}. The unit cell dimensions are $a=5.268$~\AA, $b=3.172$~\AA\ and $c=5.918$~\AA\  at ambient pressure and room temperature. Both the Mn and the P atoms occupy the $4c$ $(x, 1/4, z)$ crystallographic positions with $x_{\rm Mn} = 0.0049(2)$, $z_{\rm Mn} = 0.1965(2)$ and $x_{\rm P} = 0.1878(5)$, $z_{\rm P} = 0.5686(5)$.\cite{Rundqvist_ACS_1962_sup}

\subsubsection{Muon stopping sites}

The ab initio identification of the muon stopping site was performed with
the method described in Ref.~\onlinecite{jp5125876}.
The description of the electronic density was obtained with DFT using
a plane wave and pseudopotential approach as implemented in the
{\sc Quantum ESPRESSO} suite of codes.\cite{QE-2009}
The reciprocal space was sampled with a  $6 \times 8 \times 12$ Monkhorst-Pack
 grid.\cite{PhysRevB.13.5188}
The exchange-correlation functional of Perdew, Burke, and Ernzerhof and
the Methfessel-Paxton scheme with 0.01 Ry smearing were used.\cite{PhysRevLett.77.3865,PhysRevLett.78.1396,PhysRevB.40.3616}
The ultrasoft pseudopotentials described in Ref.~\cite{Garrity2014446}
and a basis set expanded up to a kinetic-energy cutoff of 70 Ry and up to 500 Ry for
charge density were adopted.

These settings guarantee an accurate description of the crystalline structure
 of the material.
In the collinear spin formalism, at ambient pressure, the ferromagnetic
state has the lowest enthalpy and it is therefore consider as
the ground state for the structural relaxations of the impurity in the
suprecells.

A supercell containing 129 atoms (including the muon which is described
as a hydrogen atom) is used to locate the  possible interstitial embedding positions
occupied by the muon. To get a reasonable compromise between speed and accuracy, the
kinetic energy cutoff and the charge density cutoff were reduced to 60
Ry and 400 Ry respectively. The Baldereschi point $\mathbf{k}=({1}/{4},{1}/{4},{1}/{4})$
 was used to sample the reciprocal space.\cite{PhysRevB.7.5212}
 The lattice cell parameters were kept
fixed during the relaxation.


A grid of $4 \times 4 \times 4$ initial interstitial positions  was selected to explore the
whole interstitial space of the unit cell and identify all the possible
embedding sites. After removing the positions too
close to the atoms of the hosting system (less than 1~{\AA}) and disregarding symmetry
equivalent sites, the set of 9 interstitial locations was obtained.
The structural relaxations were performed with the convergence criteria
set to $10^{-4}$ Ry for the total energy and to $10^{-3}$ Ry/a.u. for forces.
Five interstitial positions have been identified with this procedure.
The total energy differences between the possible interstitial
sites is reported in Table \ref{tab:dftmuontable}.

\begin{table}
\begin{tabular}{|c|c|c|}
\hline
Candidate muon site & Position & $\Delta E = E_{i} - E_{1}$ (eV)\\
\hline
Site 1 & (0.103, 0.25, 0.921) &  0  \\
\hline
Site 2 & (0.486, 0.935, 0.938) & 0.8  \\
\hline
Site 3 & (0.549, 0.75, 0.954) &  0.8  \\
\hline
Site 4 & (0.93, 0.25, 0.726) &  0.3  \\
\hline
Site 5 & (0.772, 0.04, 0.672) &  0.4   \\
\hline
\end{tabular}
\caption{List of the candidate muon embedding sites identified with
supercell structural relaxations. The energy differences $\Delta E$ are referred to Site 1.}\label{tab:dftmuontable}
\end{table}

The identification of multiple candidate sites is not an
unexpected feature of DFT based muon site
assignements.\cite{jp5125876,PhysRevLett.114.017602,PhysRevB.91.144417,PhysRevB.87.121108,PhysRevB.88.064423}
This is partially caused by the structural optimization algorithm which
neglects both the zero point
motion energy of the muon and the effects of temperature.
Molecular dynamics approaches would substantially improve the accuracy
of the results, but they would also result in a
tremendous increase of the computational costs.
The selected convergence criteria may also cause the relaxation
algorithm to stop in configurations which are not real minima
but rather constitute a flat area between different interstitial positions.

Since only one frequency is observed in the FM phase of MnP, it is
reasonable to assume that only one fully populated muon site is
present in this material.
Simulations performed with the double adiabatic approximation and the
exploration algorithm discussed in Ref.~\onlinecite{jp5125876} show that the
energy barrier binding the muon in site 1 is larger than 0.5 eV
while the same analysis shows that sites 4 and 5 cannot bind a muon
since their energy barriers are of the order of 0.1 eV.
Site 1 was therefore selected for the subsequent analysis of the experimental data (see Fig.~\ref{fig:MnP_Unit-cell}).

\begin{figure*}[htb]
\includegraphics[width=0.85\linewidth]{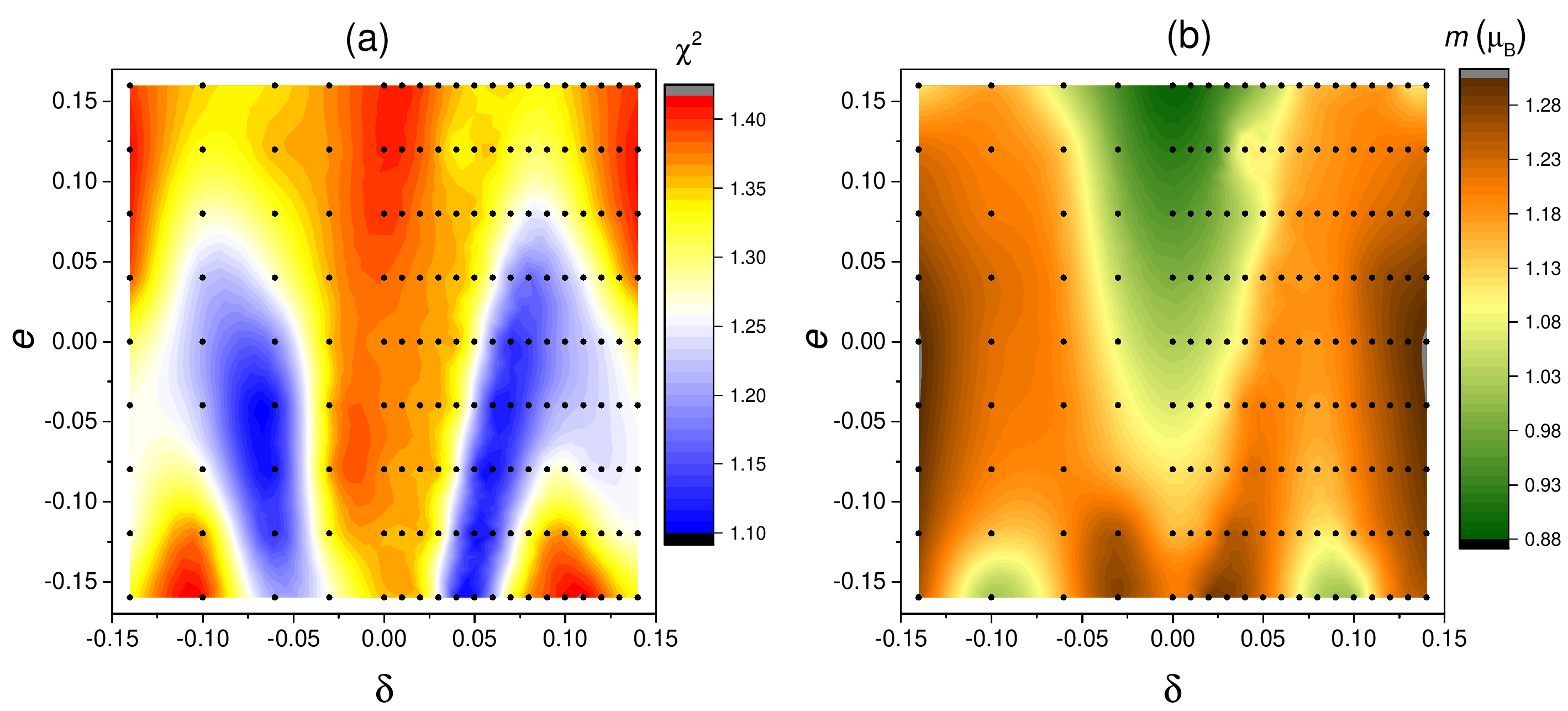}
%
\caption{Correlation plots of $\chi^2(\delta,e)$ (a) and $m(\delta,e)$ (b) as obtained from the fit of Eq.~\ref{eq:Hel-b} to the experimental data. $\delta$ is the component of the propagation vector ${\bf Q}=(0,\delta,0)$ (the positive and the negative sign correspond to the right and the left- and right-handed helix, respectively) and $e=(m_a-m_c)/(m_a+m_c)$ is the eccentricity of the elliptical helical$-b$ structure ($m_a$ and $m_c$ are components of $m$ along the $a-$ and $c-$axis, respectively). Black dots correspond to set of ($\delta$, $e$) points where the calculations and the corresponding fits were made. }
 \label{fig:Contour}
\end{figure*}

\subsubsection{Local field on the muon stopping site}

Muons probe the local field, which is the vector sum of the internal (dipolar) magnetic field and the contact field at a particular site. The spontaneous local field for the site $i$ was calculated as:
\begin{equation}
{\bf B}_{{\rm loc},i}={\bf B}_{{\rm dip},i}+{\bf B}_{{\rm cont},i}
 \label{eq:Bloc}
\end{equation}

The dipolar magnetic field $B_{\rm dip}({\bf r})$ at position ${\bf r}$ within the lattice unit cell is:\cite{Blundell_PhysB_2009_sup}
\begin{equation}
B_{\rm dip}^\alpha({\bf r})=\frac{\mu_0}{4\pi} \sum_{i,\beta} \frac{m_i^\beta}{R_i^3}
 \left( \frac{3 R_i^\alpha R_i^\beta}{R_i^2} - \delta^{\alpha \beta} \right)
 \label{eq:B_dip}
\end{equation}
Here ${\bf R}_i={\bf r}-{\bf r}_i$, $\alpha$ and $\beta$ denote the vector components $x$, $y$, and $z$, ${\bf r}_i$ is the position of $i-$th magnetic ion in the unit cell, and $m_i^\beta$ is the corresponding dipolar moment. The summation is taken over a sufficiently large Lorentz sphere of radius $R_L$.

The contact field ${\bf B}_{\rm cont}$ was obtained as:
\begin{equation}
{\bf B}_{\rm cont}=A_{\rm cont}\sum_{i=1}^N \omega(i){\bf m}_i,
 \label{eq:Bcont}
\end{equation}
where $A_{\rm cont}$ is the coupling contact constant, ${\bf m}_i$ are $N$ nearest neighboring magnetic moments and $\omega(i)$ is weight obtained as $\omega(j)=R_j^{-3}/\sum_{i=1}^N R_i^{-3}$ with $R_j$ being distance between the muon and $j$-th magnetic moment.

The contact field $ B_{\rm cont}$ was calculated by using Eq.~\ref{eq:Bcont} by considering 3 nearest neighbours for each particular muon site.

\subsubsection{Comparison of the HP-LT magnetic phase with the helical$-c$ structure from Ref.~\onlinecite{Wang_Arxiv_2015_sup} and collinear AFM structures from Ref.~\onlinecite{Gercsi_PRB_2010_sup} }

The calculations of the local fields on the muon sites were started from the known ferromagnetic and double helical magnetic structures at ambient pressure (see Refs.~\onlinecite{Huber_PR_1964_sup, Felcher_JAP_1966_sup, Obara_JPSJ_1980_sup, Forsyth_PPS_1966_sup}). This allows us to determine the coupling contact constant $A_{\rm cont}\simeq -0.447$~${\rm T}/\mu_{\rm B}$. Using the so determined $A_{\rm cont}$ and assuming that the relative positions of Mn atoms and muons  are independent on pressure, the pressure dependence of Mn moment in the FM state (Fig.~\ref{fig:Moment_vs_pressure} in the main text) was obtained. The pressure dependence of the lattice constants $a$, $b$, and $c$ was taken from Ref.~\onlinecite{Wang_Arxiv_2015_sup}.

Calculations of the local fields at the muon sites for HP helical$-c$ structure with the propagation vector ${\bf Q}=(0,0,0.25+\delta)$, suggested in Ref.~\onlinecite{Wang_Arxiv_2015_sup}, were performed by assuming that the coupling contact constant $A_{\rm cont}$ and the relative positions of Mn atoms and muons  are independent on pressure. With $m=1.17$~$\mu_{\rm B}$ per Mn atom two pairs of helical distributions
with $B_{min}$/$B_{max}$ field 0.194/0.783~T and 0.236/0.587~T were obtained, which is inconsistent with the single-peak field distribution presented in Fig.~\ref{fig:Phase-Diagramm}~d in the main text.

Gercsi and Sandeman Ref.~\onlinecite{Gercsi_PRB_2010_sup} have shown that 3 collinear antiferromagnetic structures could be realized in MnP.   The results of the calculations are summarized in Table~\ref{Table:AFMs}. None of the structures result in $B_{\rm loc}$ values comparable with the experimentally determined  $B_{int}({\rm 5~K})\simeq 0.827$~T.

\begin{table}[htb]
\caption[~]{\label{Table:AFMs} Possible collinear antiferromagnetic configurations of MnP from Ref.~\onlinecite{Gercsi_PRB_2010_sup}. Mn spins are aligned along crystallographic $b-$axis. $B_{\rm loc}$ is the calculated local field on muon sites. $A_{\rm cont}=-0.447$~$\mu_{\rm B}/{\rm T}$ is fixed from the analysis of ambient pressure data. }
\begin{center}
 \vspace{-0.5cm}
\begin{tabular}{ccccccccccc}\\
 \hline
 \hline
& Mn1& Mn2& Mn3& Mn4 & $B_{\rm loc}$\\
&&&&&(T)\\
\hline
AFM1&$\Rightarrow$&$\Rightarrow$&$\Leftarrow$&$\Leftarrow$& 0.461 \\
AFM2&$\Rightarrow$&$\Leftarrow$&$\Leftarrow$&$\Rightarrow$& 0.306 \\
AFM3&$\Rightarrow$&$\Leftarrow$&$\Rightarrow$&$\Leftarrow$& 0.398\\

 \hline \hline \\
\end{tabular}
   \end{center}
\end{table}

\subsubsection{Consistency of the HP-LT magnetic phase with helical$-b$ structure from Ref.~\onlinecite{Matsuda_Arxiv_2016_sup}}

\begin{figure}[htb]
\includegraphics[width=0.85\linewidth]{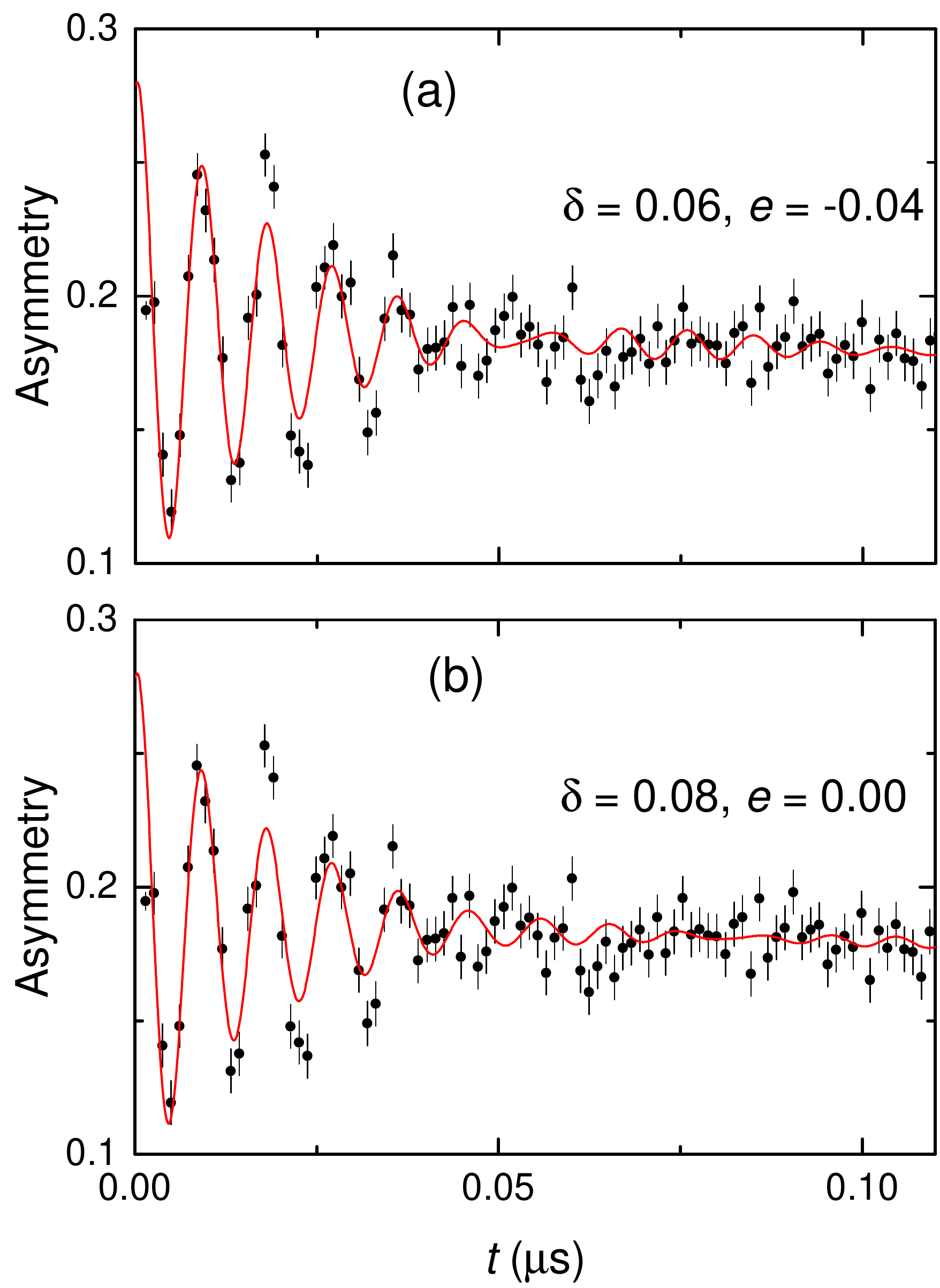}
%
\caption{ZF $\mu$SR time spectra at $p=2.42$~GPa. For increasing accuracy, two time-spectra taken at $T=5$ and 25~K were summed together. The solid lines are fits within the helical$-b$ structure for $\delta=0.6$, $e=-0.04$ (panel a) and $\delta=0.08$, $e=0.0$ (panel b). }
 \label{fig:helical-b_time-spectra}
\end{figure}

Following Ref.~\onlinecite{Matsuda_Arxiv_2016_sup} in the helical$-b$ structure the Mn moments are coupled in pairs (Mn1/Mn2 and Mn3/Mn4), and rotate within the $ac-$plane with a constant phase difference between the different pairs along ${\bf Q}=(0,\delta,0)$. The magnetic moment was found to be elongated along the crystallographic $a-$axis. At $p\simeq 1.8$~GPa and $T=6$~K, $\delta\simeq 0.09$ and the elongation of the moment $e=(m_a-m_c)/(m_a+m_c)$ is $0.15(8)$.\cite{Matsuda_Arxiv_2016_sup}

The test of the helical$-b$ structure was started form the calculations of $B_{min}$ and $B_{max}$ fields for each particular muon site in the range of $-0.14\leq \delta \leq 0.14$ and $-0.16\leq e \leq 0.16$ (black dots in Fig.~\ref{fig:Contour}). With such determined sets of $B_{min}$ and $B_{max}$ the oscillatory part of the muon-asymmetry spectra was fitted to:
\begin{eqnarray}
A_{osc}(t)&=& A_{osc}(0) \sum_{i=1}^4 \omega_i e^{-\sigma_i^2 t^2/2}  \label{eq:Hel-b} \\
&&\times J_0(\gamma_\mu \Delta B_i t)\cos(\gamma_\mu B_{av,i}t). \nonumber
\end{eqnarray}
Here $i$ denotes the $i-$th muon site, $\sigma_i$ is a Gaussian relaxation rate, $\omega_i$ is the weight ($\omega_i=0.25$ in our case since all four muon sites are equivalent), $\Delta B_i = C*(B_{max,i}-B_{min,i})/2$, and $B_{av,i}=C*(B_{max,i}+B_{min,i})/2$. The parameter $C$ accounts for the possible deviation of the magnetic moment $m$ from $m=1.17$~$\mu_{\rm B}$ as determined from calculations in the FM state (see Fig.~\ref{fig:Moment_vs_pressure} in the main text). Note that according to Eqs.~\ref{eq:Bloc}, \ref{eq:B_dip}, and \ref{eq:Bcont} the local field at the muon stoping site is directly proportional to $m$.

The goodness of fits were checked by using the $\chi^2$ criteria. As follows from Fig.~\ref{fig:Contour}~a for certain values of $\delta$ and $e$ the helical$-b$ structure becomes consistent with the experimental data. The  magnetic moment as a function of $\delta$ and $e$ is presented in Fig.~\ref{fig:Contour}~b.  The $\mu$SR time-spectra and fits within the helical$-b$ model for two representative sets of $\delta$ and $e$ parameters are shown in Fig.~\ref{fig:helical-b_time-spectra}.

\end{document}